\documentclass[aps, prb, reprint, superscriptaddress]{revtex4-1}

\usepackage{graphicx}
\usepackage{dcolumn}
\usepackage{bm}
\usepackage{amsmath}
\usepackage{amssymb}
\usepackage{upgreek}
\usepackage{textcomp}

\begin{document}

\title{The mechanism of spin-orbit coupling in a 2D oxide interface}

\author{Patrick Seiler}
\email{patrick.seiler@physik.uni-augsburg.de} 
\affiliation{Center for Electronic Correlations and Magnetism, EP VI, Institute of Physics, University of Augsburg, 86135 Augsburg, Germany}

\author{Jone Zabaleta}
\affiliation{Max Planck Institute for Solid State Research, Heisenbergstra{\ss}e 1, 70569 Stuttgart, Germany}

\author{Robin Wanke}
\affiliation{Max Planck Institute for Solid State Research, Heisenbergstra{\ss}e 1, 70569 Stuttgart, Germany}

\author{Jochen Mannhart}
\affiliation{Max Planck Institute for Solid State Research, Heisenbergstra{\ss}e 1, 70569 Stuttgart, Germany}

\author{Thilo Kopp}
\affiliation{Center for Electronic Correlations and Magnetism, EP VI, Institute of Physics, University of Augsburg, 86135 Augsburg, Germany}

\author{Daniel Braak}
\affiliation{Center for Electronic Correlations and Magnetism, EP VI, Institute of Physics, University of Augsburg, 86135 Augsburg, Germany}

\date{\today}


\begin{abstract}
The presence of spin-orbit coupling drives the anomalous magnetotransport at oxide interfaces and forms the basis for numerous intriguing properties of these 2D electron systems, such as topologically protected phases or anti-localization. 
For many of those systems, the identification of the underlying coupling mechanism is obfuscated by multi-band effects. We therefore analyze the transport of LaAlO$_3$/SrTiO$_3$ interfaces under high pressures, a technique to single out the multi-band contributions. We argue that the observed magnetoresistance is due to quantum interference and not related to Coulomb interaction. Therefore, this system is an excellent candidate to generate a metal-insulator transition of the long-sought symplectic 2D universality class. It is shown that the spin-orbit coupling can be linked unambiguously to the band structure with a cubic (Dresselhaus-like) rather than a linear (Rashba-like) spin-orbit band splitting. 
\end{abstract}



\maketitle


\section{Introduction}
\label{sec:Intro}

Research on oxide heterostructures has been very fruitful during the last decade, in particular due to the special electronic properties of interface layers between oxides slabs of disparate stoichiometry. These properties are determined by electronic reconstruction, that is, by the formation of interface states which differ significantly from the electronic states of the adjacent bulk materials. In the paradigmatic LaAlO$_3$/SrTiO$_3$ oxide heterostructure,~\cite{Ohtomo2004} two band insulators share a metallic interface when at least four layers of LaAlO$_3$ (LAO) with LaO termination towards the TiO$_2$ interface layer are deposited on SrTiO$_3$ (STO).~\cite{Thiel2006} However, electronic reconstruction not only allows for a conducting state but, moreover, interface confined superconducting,~\cite{Reyren2007} magnetic,~\cite{Brinkman2007, Li2011a, Bert2011, Ruhman2014} and 
negative compressibility states~\cite{Li2011b,Tinkl2012} have been identified. Even topologically protected superconducting phases have been predicted to form at these heterostructures.~\cite{Fidkowski2013,Scheurer2015,Loder2015}

Beyond the fascinating fundamental physical problems, related  to the build-up and the coexistence or competition of these interface states, the oxide heterostructures are also in the focus of research with regard to their functionality, that is, their potential to use them as electronic devices.~\cite{Mannhart2010} For example, circuits with all-oxide field-effect transistors were fabricated from LAO/STO heterostructures.~\cite{Foerg2012} 
The wider use of such oxide devices requires to characterize the electronic transport under the effect of external parameters such as temperature, magnetic field, pressure and bias. However, the electronic transport at the interface of LAO/STO is complex:  it is confined to few layers of STO with conduction electrons from 
all three Ti $t_{2g}$ states, so a multi-band quasi two-dimensional (2D) behavior is to be explored. Moreover, spin-orbit coupling (SOC) seems to play an important role in these systems, affecting the electron propagation in the proximity to the disordered interface.~\cite{Caviglia2010} 
Disorder may be generated by impurities, such as oxygen vacancies, or by intermixing and structural defects. Eventually, also electron interaction effects are of relevance.~\cite{Breitschaft2010, Fuchs2015} A metal-insulator transition (MIT) depending on charge carrier density has been observed at low densities 
but its origin has not been clarified.~\cite{Liao2011}

In this intricate context, fundamental issues are not yet adequately understood. Several of them are related to transport in disordered electronic systems: The strong SOC, which can be tuned via a gate voltage,~\cite{Caviglia2010} is assumed to lead to weak Anderson anti-localization (WAL), but until now there has been no clear picture of the specific associated mechanism: This WAL may originate either from elastic potential scattering in combination with Rashba SOC for the propagating electrons or from impurities with spin-orbit coupling. While this quantum interference phenomenon is in its very nature anti-localizing, renormalization group arguments within a one-parameter scaling theory predict an Anderson metal-insulator transition in 2D, belonging to the symplectic universality class.~\cite{Wegner1989} We therefore assess that WAL, specifically at LAO/STO interfaces, may be viewed as a precursor of this transition, which has never been observed experimentally up to now.   

However, the observed magnetoconductance may as well be attributed to electron-electron-interaction (EEI),~\cite{Fuchs2015} and we readdress 
this controversy concerning its origin.
It turns out that the multi-band character of the interface plays a key role in
resolving the issue.

We investigate magnetoresistance (MR) measurements at the LAO/STO interface under hydrostatic pressure and find that the WAL becomes visible only for sufficient pressure in our samples, which feature a relatively high carrier density $>$6\,$\cdot$\,10$^{13}$\,cm$^{-2}$. We show that a careful analysis of the contribution of multiple charge carriers and their interplay with WAL may resolve previous puzzles of the low temperature behaviour.
Up to now, the overlying multi-band nature of the electronic interface state made it challenging to favor a certain SOC scenario for the WAL.\cite{Caviglia2010, Hernandez2012, Stornaiuolo2014} We introduce a fitting procedure that allows for the evaluation of the magnetotransport data over the complete measured magnetic field range. This fitting procedure treats the multiple charge carrier Hall effect and the WAL on the same footing. A main result from this analysis is the clear evidence that WAL arises due to cubic spin-orbit splitting of interface electronic bands.

This paper is structured as follows: In Section~\ref{sec:WAL}, we review the model for the itinerant $t_{2g}$ bands of LAO/STO and the formulation of WAL theory, depending on the microscopic structure of the spin-orbit coupling. In Section~\ref{sec:exp}, we present the treatment of the multi-band Hall effect when a significant WAL is present and provide the experimental evidence for the cubic spin-orbit splitting in our samples. This links the WAL to the corresponding 
Ti $d_{{xz}}$/$d_{{yz}}$ bands. In Section~\ref{sec:disc} we discuss the role of interaction effects. Finally, we consider the possibility to observe a symplectic MIT in LAO/STO heterostructures.


\section{Weak anti-localization}
\label{sec:WAL}

The electron liquid at the LAO/STO interface is strongly confined to a few lattice constants in the transverse direction.~\cite{Sing2009, Delugas2011} As a two-dimensional system it cannot support a simple metallic state: assuming that electron interactions are irrelevant, any disorder will lead to localization of the electronic system, at least for low temperatures.\cite{Abrahams1979} However, the strong spin-orbit coupling present in this system opens the possibility of anti-localization being visible in the magnetotransport.\cite{Hikami1980} 
Both localization and anti-localization are interference effects of the electron wave function and can be detected via sheet resistance measurements in transverse magnetic fields. The field-dependence of the magnetoresistance is controlled by the type of spin-orbit coupling.


\subsection{Model for spin-orbit coupling in LaAlO$_3$/SrTiO$_3$ heterostructures}
\label{subsec:SOCLAOSTO}

\begin{figure}
    \includegraphics[width=\columnwidth]{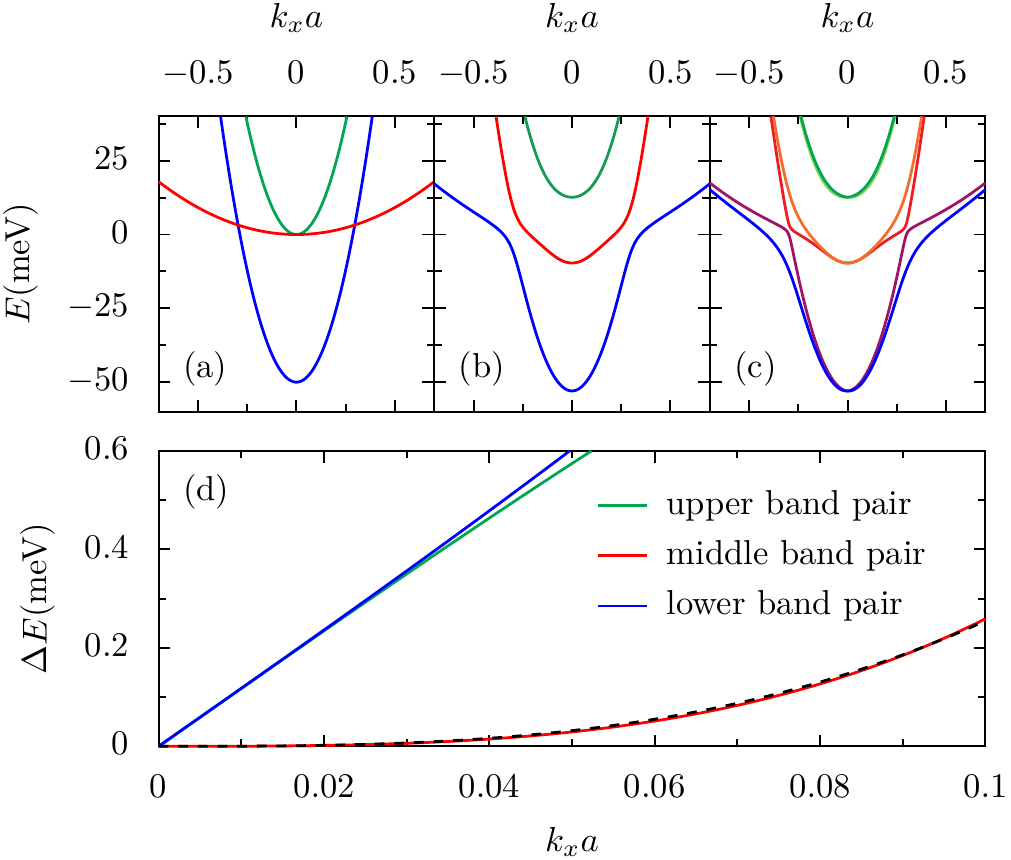}
    \caption{\label{Fig:bands} \emph{Energy dispersion in the six-band model of {\rm LAO/STO}.} Band structure for $m_\text{h}=6.8\,m_\text{e}$, $m_l=0.41\,m_\text{e}$, $\Delta_z=50\,\text{meV}$ and (a) $\Delta_\text{aso}=\Delta_\text{if}=0$; (b) $\Delta_\text{aso}=9.65\,\text{meV}$, $\Delta_\text{if}=0$; (c) $\Delta_\text{aso}=9.65\,\text{meV}$, $\Delta_\text{if}=20\,\text{meV}$. The combination of atomic spin-orbit coupling and interface buckling leads to a dispersive spin-splitting, which is linear in momentum for the lower and higher band pair, but cubic for the middle band pair. (d) The spin splitting is shown for the specific band pairs, the dashed line is a cubic fit.}
\end{figure}

To analyze the experimental MR data, we base the evaluation on the Ti $3d$ $t_{2g}$ electronic structure at the interface~\cite{Pentcheva2009, Pavlenko2011, Cossu2013}
and its six-band extension introduced in Refs.~\onlinecite{Zhong2013, Khalsa2013} for the system with SOC. The corresponding tight-binding Hamiltonian contains four parts,
\begin{equation}
    \hat{\mathcal{H}} = \hat{\mathcal{H}}_0 + \hat{\mathcal{H}}_\text{aso} + \hat{\mathcal{H}}_\text{if} + \hat{\mathcal{H}}_\text{imp},
    \label{eq:LAOSTO}
\end{equation}
which we will briefly review in the following.
The electrons at the interface reside mainly in the $t_{2{g}}$ bands. Symmetry breaking at the interface induces an energy shift $\Delta_{z}$ of the $xy$-band. The kinetic part of the Hamiltonian becomes block diagonal in momentum space. A single block $h_0$, given in the $\{yz,\,zx,\,xy\}$ basis, reads
\begin{equation}
    h_0 =
    \begin{pmatrix}
        \frac{\hbar^2 k_{x}^2}{2 m_\text{h}} + \frac{\hbar^2 k_y^2}{2 m_\text{l}} & 0 & 0 \\
        0 & \frac{\hbar^2 k_{x}^2}{2 m_\text{l}} + \frac{\hbar^2 k_{y}^2}{2 m_\text{h}} & 0 \\
        0 & 0 & \frac{\hbar^2 k_{x}^2}{2 m_\text{l}} + \frac{\hbar^2 k_{y}^2}{2 m_\text{l}} - \Delta_{z}
    \end{pmatrix} \otimes \sigma_0
    \label{eq:h_0}
\end{equation}
where $\sigma_0$ is the identity in spin space. The energy shift $\Delta_{z}=50\,$meV, as measured by x-ray~\cite{Salluzzo2009} and confirmed in {\it ab initio} evaluation,~\cite{Zabaleta2016} and the band masses $m_\text{h}=6.8\,m_\text{e}$ and $m_\text{l}=0.41\,m_\text{e}$, which are the masses of heavy and light charge carriers, respectively, have been extracted from electronic structure evaluations.\cite{Zabaleta2016, Zhong2013, Khalsa2013} To take care of the atomic spin-orbit coupling (SOC) at the Ti sites, the local spin splitting
\begin{equation}
    h_\text{aso} = \Delta_\text{aso}
    \begin{pmatrix}
        0 & i \sigma_{z} & -i \sigma_{y} \\
        -i \sigma_{z} & 0 & i \sigma_{x} \\
        i \sigma_{y} & - i \sigma_{x} & 0
    \end{pmatrix}
\end{equation}
is included. The broken structure inversion symmetry at the interface is implemented through interorbital hoppings, approximately linear in momentum,
\begin{equation}
    h_\text{if} = \Delta_\text{if}
    \begin{pmatrix}
        0 & 0 & i k_{x} \\
        0 & 0 & i k_{y} \\
        -i k_{x} & - i k_{y} & 0
    \end{pmatrix}\otimes \sigma_0.
\end{equation}
The latter Hamiltonian part $h_\text{if}$ plays a crucial role for the Rashba-like effects at interfaces~\cite{Winkler2010,Zhong2013} because, combined with the local SOC, it describes an effective dispersive spin-orbit splitting, see Fig.~\ref{Fig:bands}. 

Wheras the original Rashba effect,\cite{Bychkov1984} derived from the Dirac equation in a one-band system, describes a linear spin splitting and is very small, the multi-band origin leads here to a more complex SOC structure with sizeable splitting (for a review, see Ref.~\onlinecite{Winkler2010}). Near the $\Gamma$-point, one finds a linear spin splitting with respect to $k$ for the lower and higher band pair, but a cubic spin splitting, reminiscent of Dresselhaus SOC, in the middle band pair.~\cite{Zhong2013, Khalsa2013, Kim2013} Note that this picture breaks down when the filling is close to the avoided band crossings. The Rashba-like effect is not directly affected by an external electric (gate) field, but rather tuned via the concomitant change of the band filling.~\cite{Ilani2012} Moreover, $\Delta_\text{if}$ is supposed to be field dependent.~\cite{Steffen2015} 

Finally, we introduce disorder into the Hamiltonian by $\hat{\mathcal{H}}_\text{imp}$, which represents potential scattering at dislocations that are at position $\mathbf{r}_i$,
\begin{equation}
    \hat{\mathcal{H}}_\text{imp} = \sum\limits_i V( \hat{\mathbf{r}} - \mathbf{r}_i ).
\end{equation}
The present model agrees well with band structure measurements using x-ray absorption spectroscopy and ARPES on LAO/STO.\cite{Salluzzo2009, Berner2013, Cancellieri2014}

Although earlier magnetotransport studies on LAO/STO interfaces indicated the importance of choosing the correct theoretical description,\cite{Kim2013, Nakamura2012} no consistent picture has been achieved so far. In the following, we shall review the different WAL evaluations before comparing them with experimental results in Section~\ref{sec:exp}.


\subsection{Linear vs. cubic spin-orbit splitting}

A spin-orbit coupling term
in the kinetic energy, combined with elastic scattering processes, is able to explain weak anti-localization signatures of transport measurements in two-dimensional systems.\cite{Iordanskii1994, Pikus1995, Knap1996, Hassenkam1997} The theory in its final form was established by Iordanskii, Lyanda-Geller and Pikus (ILP) and treats 2D Hamiltonians of the type
\begin{equation}
    \hat{\mathcal{H}}^\text{ILP} = \frac{ \hat{\textbf{p}}^2}{2 m} + \hbar  \boldsymbol\upsigma \cdot \boldsymbol\Omega \left( \hat{\textbf{p}} \right) + \sum\limits_i V( \hat{\mathbf{r}} - \mathbf{r}_i )
    \label{eq:ILP_Ham}
\end{equation}
where $\boldsymbol\upsigma = (\sigma_x,\sigma_y)$ 
and $2 \boldsymbol\Omega / \hbar $ represents a precession vector that results from the broken inversion symmetry. The spin-splitting of the bands is equal to $2 \hbar |\boldsymbol\Omega (\textbf{p})|$. This splitting can either be linear or cubic in momentum, and the corresponding precession vector will be denoted as $\boldsymbol\Omega^{(1)}$ or $\boldsymbol\Omega^{(3)}$, respectively.
In combination with elastic scattering, which is characterized by a time scale $\tau$, the SOC leads to a D'yakonov-Perel\cite{Dyakonov1972} spin relaxation time $\tau_\text{so} \propto \tau^{-1}$. 
In principle, both, linear and cubic splitting can be present at the same time, which leads to a complicated formula for the magnetoresistance. However, in the case of a purely linear SOC splitting (motivated by the Rashba SOC), Punnoose worked out a closed form expression for the WAL correction to the magnetoconductivity:\cite{Punnoose2006}
\begin{equation}
    \begin{split}
     \Delta \sigma (B) = \frac{e^2}{2 \pi h} \left[ 
                                            \sum\limits_{s=0, \pm 1} u_s \psi \left( \frac{1}{2} + \frac{B_\text{i}}{B} - v_s \right) \right. \\[0.2cm] \left.
                                            - \psi \left( \frac{1}{2} + \frac{B_\text{i}}{B} \right) - 2 \ln \left( \frac{B_\text{i}}{B} \right) \right. \\[0.2cm] \left.
                                            + \frac{ 4 B^2 }{ 4 \left( B_\text{so} + B_\text{i} \right)^2 - B^2 } +
                                            C
                                        \right].
    \end{split}
    \label{EqPunnoose1}
\end{equation}
Explicit formulae for $C(B_\text{so}, B_\text{i})$, $u_s(B, B_\text{so}, B_\text{i})$ and $v_s(B, B_\text{so}, B_\text{i})$ are specified in Appendix~\ref{app:Pun}; $\psi$ is the digamma function and the effective magnetic fields $B_\text{so/i}$ are given by 
\begin{equation}
    B_\text{so/i} = \frac{\hbar}{ 4 e D \tau_\text{so/i} },
\end{equation}
where $D$ is the diffusion constant, $\tau_\text{i}$ is the relaxation time for inelastic scattering processes, and $\tau_\text{so}$ is the aforementioned time scale related to the D'yakonov-Perel spin relaxation.

Otherwise, selecting a purely cubic spin-splitting (historically motivated by the Dresselhaus effect) leads to\cite{Iordanskii1994} 
\begin{align}
    \Delta \sigma(B) = \frac{e^2}{\pi h} &\left[             \Psi\left(  \frac{B}{B_\text{so} + B_\text{i}} \right) -\frac{1}{2} \Psi\left( \frac{B}{B_\text{i}} \right) \nonumber \right. \\ &\left.
                                                              + \frac{1}{2} \Psi\left( \frac{B}{2 B_\text{so}  + B_\text{i}} \right) 
                                                   \right]
    \label{eq:Hikami}
\end{align}
where 
$    \Psi (x) = \ln(x) + \psi(\tfrac{1}{2} + \tfrac{1}{x}) $.

An expression for the quantum correction to the magnetoconductivity, similar to Eq.~\eqref{eq:Hikami}, was first derived by Hikami, Larkin, and Nagaoka (HLN).\cite{Hikami1980} However, in contrast to the ILP approach, they did not consider SOC in the orbital motion but instead scattering processes with a strong local atomic SOC
\begin{equation}
    \hat{\mathcal{H}}^\text{HLN} = \frac{ \hat{\mathbf{p}}^2}{2 m} + \sum\limits_i \left( 1+ \frac{\hat{\mathbf{p}} \cdot \boldsymbol\upsigma}{4 m^2 c^2}  \times \nabla \right) V( \hat{\mathbf{r}} - \mathbf{r}_i ).
\end{equation}
These spin-orbit scatterers lead, in the absence of inversion symmetry breaking, to a Elliott-Yafet spin relaxation mechanism with $\tau_\text{so} \propto \tau$.\cite{Elliott1954, Yafet1983}
In Ref.~\onlinecite{Hikami1980} it was stressed that no exact two-dimensional material may show weak anti-localization due to the latter scattering mechanism, because its origin lies in the $\sigma_{x}$ and $\sigma_{y}$ channels, whereas only the scattering rate for the $\sigma_{z}$ channel is finite in HLN: $\tau_{z}^{-1} \propto |\mathbf{k} \times \mathbf{k}^\prime|_{z}^2$. Therefore, the original approach was adequate to explain MR data in thin metallic films with non-vanishing scattering along the $z$-direction, but is not appropriate for truly two-dimensional systems as they appear in semiconductor quantum wells and heterostructures.\footnote{From a technical viewpoint, the HLN formula may be used for an interpretation in terms of the Dy'akonov-Perel spin relaxation mechanism\cite{Dyakonov1972, Altshuler1981, Dresselhaus1992}, recovering Eq.~\eqref{eq:Hikami}. This seems to be the reason why the HLN treatment led so far to good agreement with magnetotransport data on LAO/STO heterostructures\cite{Caviglia2010, Stornaiuolo2014, Hernandez2012}. Yet the HLN approach corresponds to qualitatively different physics than the ILP approach, and we apply ILP because the
contribution from spin-orbit scatterers is supposed to be negligible.}

Following Ref.~\onlinecite{Kim2013}, we argue that for cubic spin splitting in the LAO/STO Hamiltonian, 
Eq.~\eqref{eq:Hikami} is to be used for fitting experimental data, whereas Eq.~\eqref{EqPunnoose1} is appropriate for linear spin splitting. Therefore,  magnetotransport measurements are a means to reveal whether the observed WAL is related to the $d_{xz}$/$d_{yz}$ bands or to the $d_{xy}$ band.
However, due to the complicated multi-band nature of the heterostructure, the contribution from WAL for higher fields is obfuscated by other effects that have to be disentangled first before a meaningful fit of the WAL correction can be done.   


\section{Weak anti-localization for multi-band systems}
\label{sec:exp}

\subsection{Experimental}

The 2D electron system at the LAO/STO interface measured in this work was formed in Hall-bar patterned samples with 6 unit cells of LAO grown by Pulsed Laser Deposition onto TiO$_2$ terminated STO substrates. For applying hydrostatic pressure, samples were cut and thinned down to 1\,$\times$\,1\,$\times$\,0.2\,mm$^{3}$ to fit into a commercial piston cylinder cell (Almax Easy Lab). Details of the sample growth and of the experimental setup are available in Ref. \onlinecite{Zabaleta2016}. Two different samples, sample A and sample B, grown identically, were pressurized and measured in separate experimental runs. At every pressure step reported here the sheet resistance $R_\text{S}$ and the Hall resistance $R_\text{H}$ were recorded in 4-wire Hall bar configuration as a function of magnetic field (maximum sweeps between $\pm$8T) at various temperatures down to 2\,K using a Physical Properties Measurement System (Quantum Design). 

\begin{figure}
    \includegraphics[width=\columnwidth]{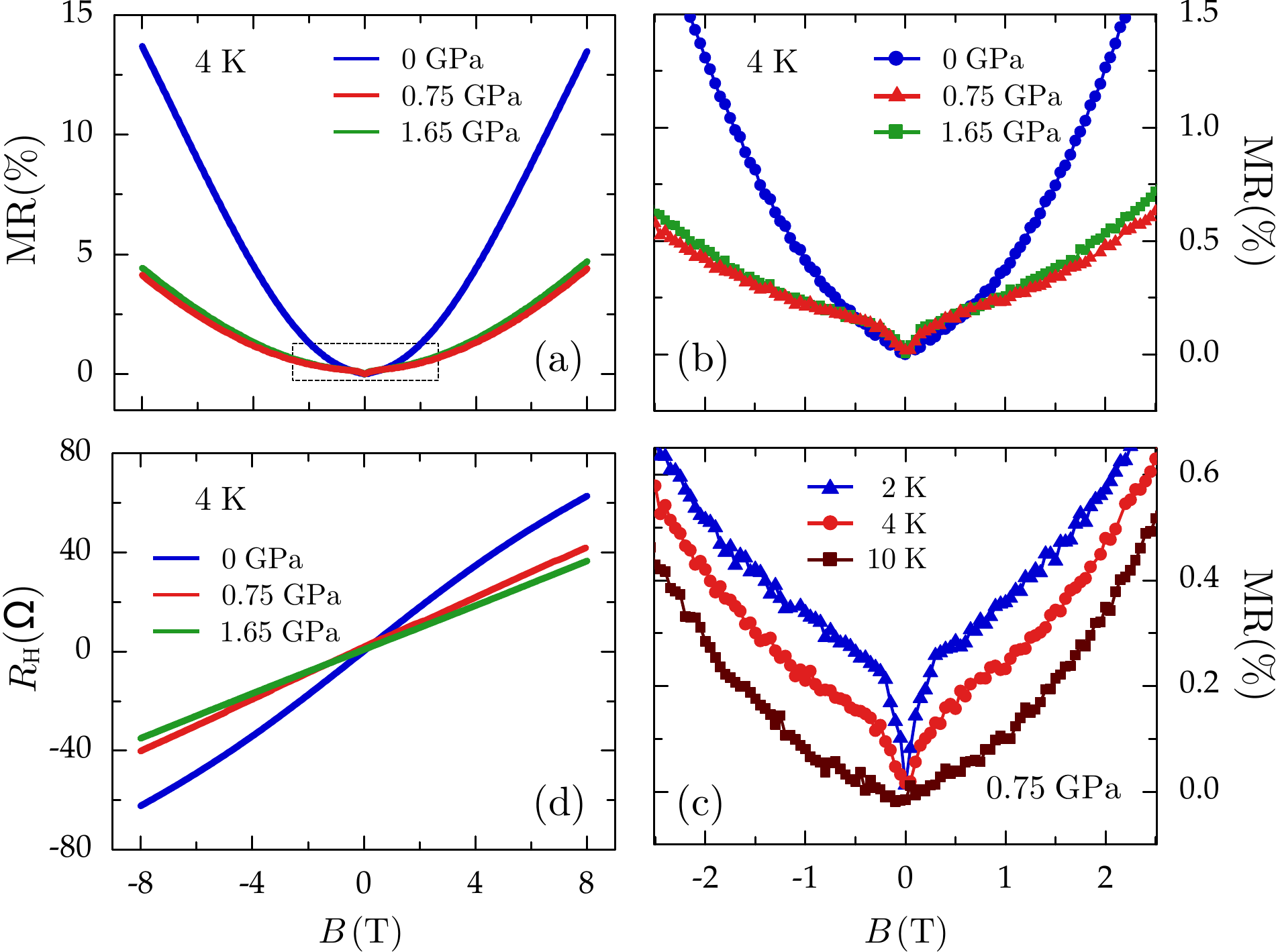}
    \caption{ \label{Fig:SampleA_MRRH} \emph{Experimental magnetotransport results for sample A.} (a) Magnetoresistance (MR) measured at 4\,K at three different pressure values. The dashed square indicates the magnified area plotted in (b). (c) MR data at 0.75\,GPa for different temperatures. The dip at zero field is enhanced with decreasing $T$. (d) Hall resistance ($R_\text{H}$) as a function of magnetic field showing the pressure-induced suppression of the multi-band signature.}
\end{figure}

Figure~\ref{Fig:SampleA_MRRH}(a) and (b) show the experimental magnetotransport data for sample A at 4\,K at three different pressure values for the magnetic field applied perpendicular to the LAO/STO interface. At ambient pressure (0\,GPa) the magnetoresistance (MR) shows the characteristic positive curvature of a multi-band system [Fig.~\ref{Fig:SampleA_MRRH}(a)].~\cite{Jost2015, Seo2009, Shalom2010, Lerer2011, Rakhmilevitch2013, Guduru2013a, Guduru2013b, McCollam2014, Joshua2013, Dikin2011} Interestingly, pressure strongly reduces this positive MR from around 14\,$\%$ at $\pm$8\,T and ambient pressure to below 5\,$\%$ under pressure. The MR shows a dip at zero magnetic field together with a slight change in the curvature at low fields, better appreciated in the zoomed-in area shown in Fig.~\ref{Fig:SampleA_MRRH}(b). It is also noteworthy that the reduction of positive MR does not continue with increasing pressure all the way but it seems to have reached a maximum somewhere around 0.75\,GPa before decreasing again at 1.65\,GPa, the maximum applied pressure. These results are entirely reproduced by sample B (see Fig.~\ref{Fig:MRRH_SampleB} in the Supporting Experimental Material, Appendix ~\ref{app:sup}), where additional intermediate pressure steps were also measured. The dip at zero-field is more pronounced the lower the temperature is and it washes away already at 10\,K, as shown in Fig.~\ref{Fig:SampleA_MRRH}(c). The suppression of the multi-band character seen in the MR data is also apparent in the Hall resistance data [Fig.~\ref{Fig:SampleA_MRRH}(d)]: the s-shape curve measured at 0\,GPa becomes increasingly linear with pressure, a feature reproduced by sample B (Fig.~\ref{Fig:MRRH_SampleB}(b) in App.~\ref{app:sup}). As a technical note, Fig.~\ref {Fig:SampleA_MRRH} shows the raw magnetotransport data, whereas for the fitting procedure described later we have symmetrized the sheet resistance data to exclude transverse components, and anti-symmetrized the Hall resistance data to get rid of longitudinal components. One can find the resulting plots in Fig.~\ref{Fig:MRRH_SampleASymm} of App.~\ref{app:sup}.
 

\subsection{Coupling of WAL and multi-band Hall effect}

In the simplest approximation, the standard textbook two-band Hall effect already leads to a nontrivial magnetoresistance response and a non-linear Hall signal.\cite{Ashcroft1976} Here one has to treat the combination of multiple band Hall effect and  WAL in a careful manner. The formulae for the full resistivity tensor in the presence of WAL and the two-band Hall effect are given in Appendix~\ref{app:Coup}.

Our evaluation is consistent with the asumption that WAL contributes only from one of the bands, to which we refer in the following as type-1 charge carriers.
The magnetoresponse, up to cubic order in the magnetic field, is given by
\begin{align}
        \text{MR} &= \dfrac{R_\text{S} (B) - R_\text{S} (0)}{R_\text{S} (0)} = a_0 + a_2 B^2, \label{Eq:MRfull} \\[0.3cm]
        R_\text{H} &= a_1 B + a_3 B^3
        + a_\text{c} B, \label{Eq:Hallfull}
\end{align}
where $R_\text{S}$ is the sheet resistance and 
\begin{align}
    a_0 &= \dfrac{1}{1+ \dfrac{\Delta \sigma}{ e \left( n_1 \mu_1 + n_2 \mu_2 \right) } } -1, \label{eq:2b1}\\[0.3cm]
    a_1 &= \dfrac{ n_1 \mu_1^2 + n_2 \mu_2^2 }{ e \left( n_1 \mu_1 + n_2 \mu_2 \right)^2 }, \\[0.3cm]
    a_2 &= \dfrac{n_1 \mu_1 n_2 \mu_2 \left( \mu_1 -\mu_2 \right)^2}{ \left( n_1 \mu_1 + n_2 \mu_2 \right)^2 }, \\[0.3cm]
    a_3 &= - \dfrac{ n_1 n_2 \mu_1^2 \mu_2^2 \left( n_1 + n_2 \right) \left( \mu_1 - \mu_2 \right)^2 }{ e \left( n_1 \mu_1 + n_2 \mu_2 \right)^4,   }\\[0.3cm]
    a_\text{c} &= \dfrac{ 2 \left( \mu_1 - \mu_2 \right) n_2 \mu_2 \Delta \sigma }{ e^2 \left( n_1 \mu_1 + n_2 \mu_2 \right)^3 },\label{eq:2b5}
\end{align}
and $n_\alpha$ and $\mu_\alpha$ are the sheet carrier densities and the mobilities of electrons of the respective bands.
Note that the parameters $a_\text{c}$ and $a_0$ control the impact of WAL from Eq.~(\ref{EqPunnoose1}) or alternatively Eq.~(\ref{eq:Hikami})---whereas $a_1$, $a_2$ and $a_3$ depend only indirectly on the WAL corrections through the sheet carrier densities and mobilities. We use Eqs.~\eqref{eq:2b1}-\eqref{eq:2b5} to describe transport in the LAO/STO interface with an effective two-band model.

In the literature, magnetotransport data have been occasionally evaluated by fitting and subsequent substraction of the background in the high field regime.
However, this approach may lead to biased conclusions as it favors the cubic spin splitting scenario: The amplitude of the WAL signal for higher fields 
is negligible in the cubic case, whereas in the linear splitting scenario the WAL signal may still yield a significant contribution. 

\begin{figure}
    \includegraphics[width=\columnwidth]{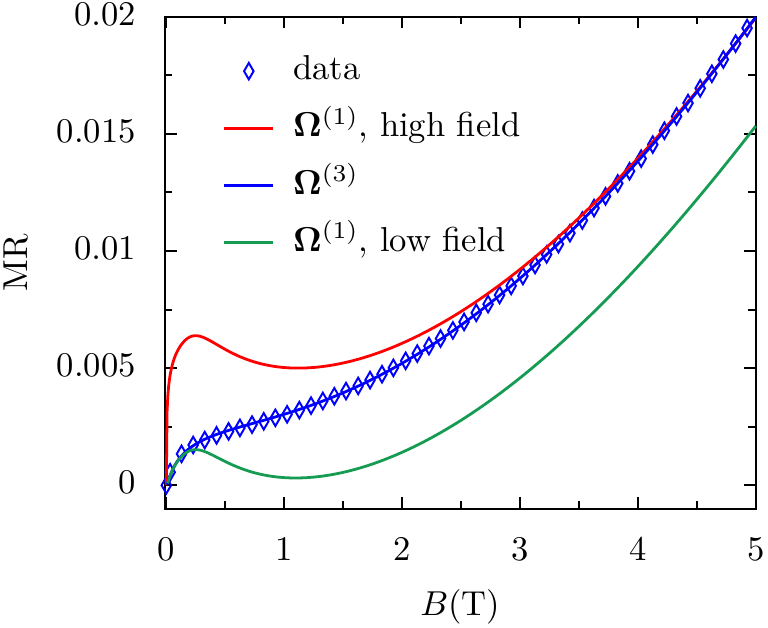}
    \caption{ \label{Fig:Fitcomp} \emph{Comparison of linear and cubic SOC fits to experimental data.} We show the results for sample B at 2\,K and 0.80\,GPa. Cubic SOC $\boldsymbol\Omega^{(3)}$, using Eq.~\eqref{eq:Hikami}, is in good agreement with the experimental data for $B_\text{i}=6\,\text{mT}$  and $B_\text{so}=0.2\,\text{T}$ (blue). For linear SOC $\boldsymbol\Omega^{(1)}$ (with Eq.~\eqref{EqPunnoose1}), the same parameters lead to good agreement only for low fields (green). With $B_\text{so}=0.2\,\text{T}$ fixed and $B_\text{i}=0.35\,\text{mT}$ optimized, the high field behaviour can be fitted, but it overshoots in the low field regime (red). If both values are optimized for the high field regime, the $B_\text{so}$ becomes unphysically large, and the structure for low fields is no longer captured. For the linear SOC, no choice of parameters allows to fit the data for all measured fields.}
\end{figure}

We numerically fit the WAL and multi-band contributions within self-consistent iterations. First, the coupling term $a_\text{c} \propto \Delta \sigma B$ is neglected and we extract $a_1$, $a_2$, and $a_3$ as well as the WAL parameters, $B_\text{i}$ and $B_\text{so}$, self-consistently. In the next step, the WAL parameters are used to fine-tune $a_3$ in a second self-consistent loop. In this procedure, fits using Eq.~\eqref{eq:Hikami} lead to very good results, similar to previous reports for STO surface measurements.~\cite{Nakamura2012} Conversely, we find that no meaningful fit can be generated using Eq.~\eqref{EqPunnoose1}. A comparison of the results for the model with linear and that with cubic spin-orbit splitting is shown in Fig.~\ref{Fig:Fitcomp}. This leads to an unambiguous identification of the $d_{xz}$/$d_{yz}$ band pair being responsible for 
the WAL signature in our system.  

\begin{figure}
    \includegraphics[width=\columnwidth]{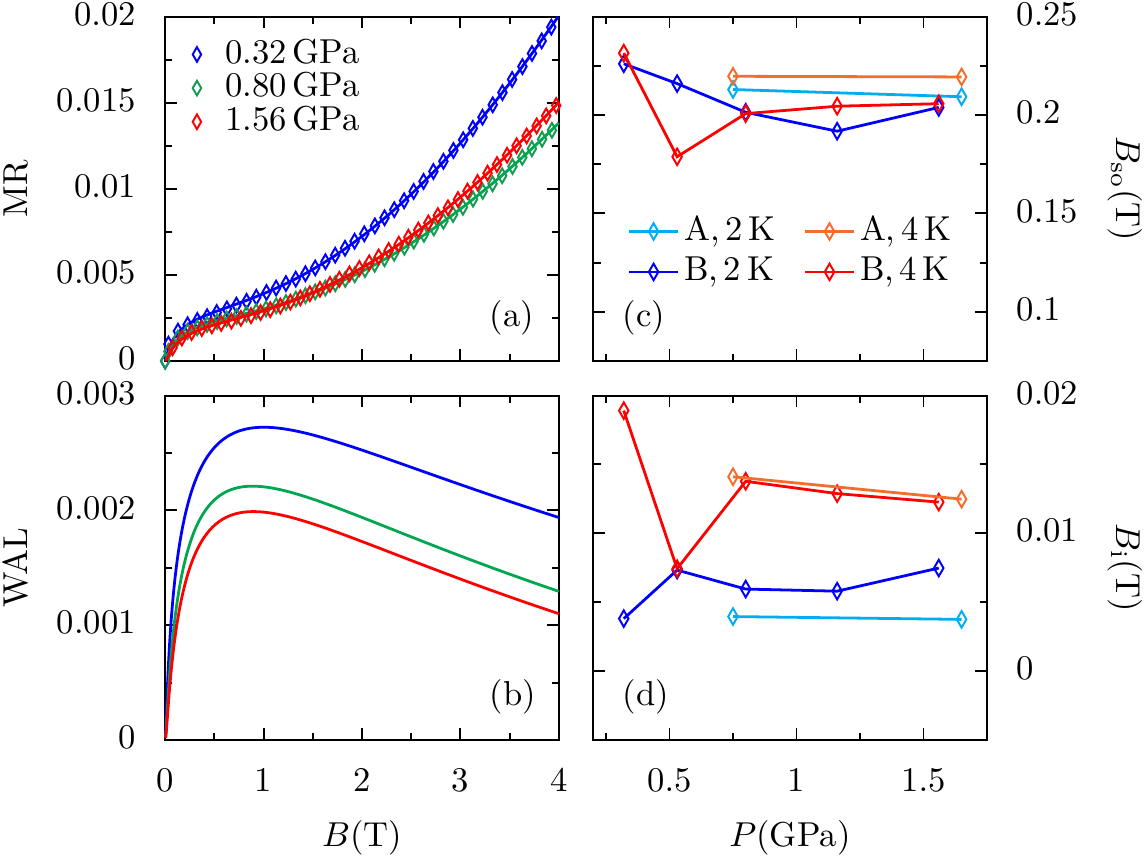}
    \caption{ \label{Fig:Fit} \emph{Results under hydrostatic pressure.} (a) Fits for sample B at 2\,K. The data for different pressures can be well fitted using~Eqs.~\eqref{Eq:MRfull} and~\eqref{Eq:Hallfull}. Curves in (b) are the
$\boldsymbol\Omega^{(3)}$ fits.
(c) and (d) show the WAL parameters for both samples, lines are a guide to the eye. They are very similar and do not depend much on pressure.}
\end{figure}

The results of the self-consistent procedure are shown in Fig.~\ref{Fig:Fit}.
We find that the parameters are similar in both of our samples. The effective spin-orbit field $B_\text{so}\approx 0.2$\,T for 2\,K as well as for 4\,K. Also, the effective inelastic field $B_\text{i}\approx 0.005$\,T for 2\,K and $\approx 0.015$\,T for 4\,K. The values are remarkably stable under pressure, although the band parameters show a strong pressure dependence. 
The inelastic scattering rate grows with temperature, in agreement with the temperature dependence of $B_\text{i}$.


\subsection{Multi-band results under pressure}

Surprisingly, in the majority of Hall signals we find a positive $a_1$ as well as a positive $a_3$, giving strong evidence for transport by high mobility electron-like and high-density hole-like charge carriers. This feature survived in the fitting procedure with and without consideration of the term~$a_\text{c}$, and cannot be explained by taking more than two electron-like charge carrier types into account. We conclude that this hole band is a stable result within the framework of our approach, but any deeper explanation goes beyond the scope of this paper, especially because the theoretically predicted hole band on the surface of the LAO layer is usually not found in experiments~\cite{Berner2013} without special preparation such as capping layers.\cite{Pentcheva2010} We assume that the WAL arises from the electron-like carriers (type-1) only. In fact, this interpretation is the only one consistent with the band structure obtained from {\it ab initio} calculations, which contain only electron-like bands with SOC. For our purpose the detailed nature of the second band of carriers is rather unimportant: The explanation of WAL requires only disentangling the (electron-like) band with strong spin-orbit coupling from the contribution of the second band in the magnetotransport measurements. Nevertheless, it is interesting to note that these measurements alone are sufficient to identify a second band including carrier type, mobility and density.

\begin{figure}
    \includegraphics[width=\columnwidth]{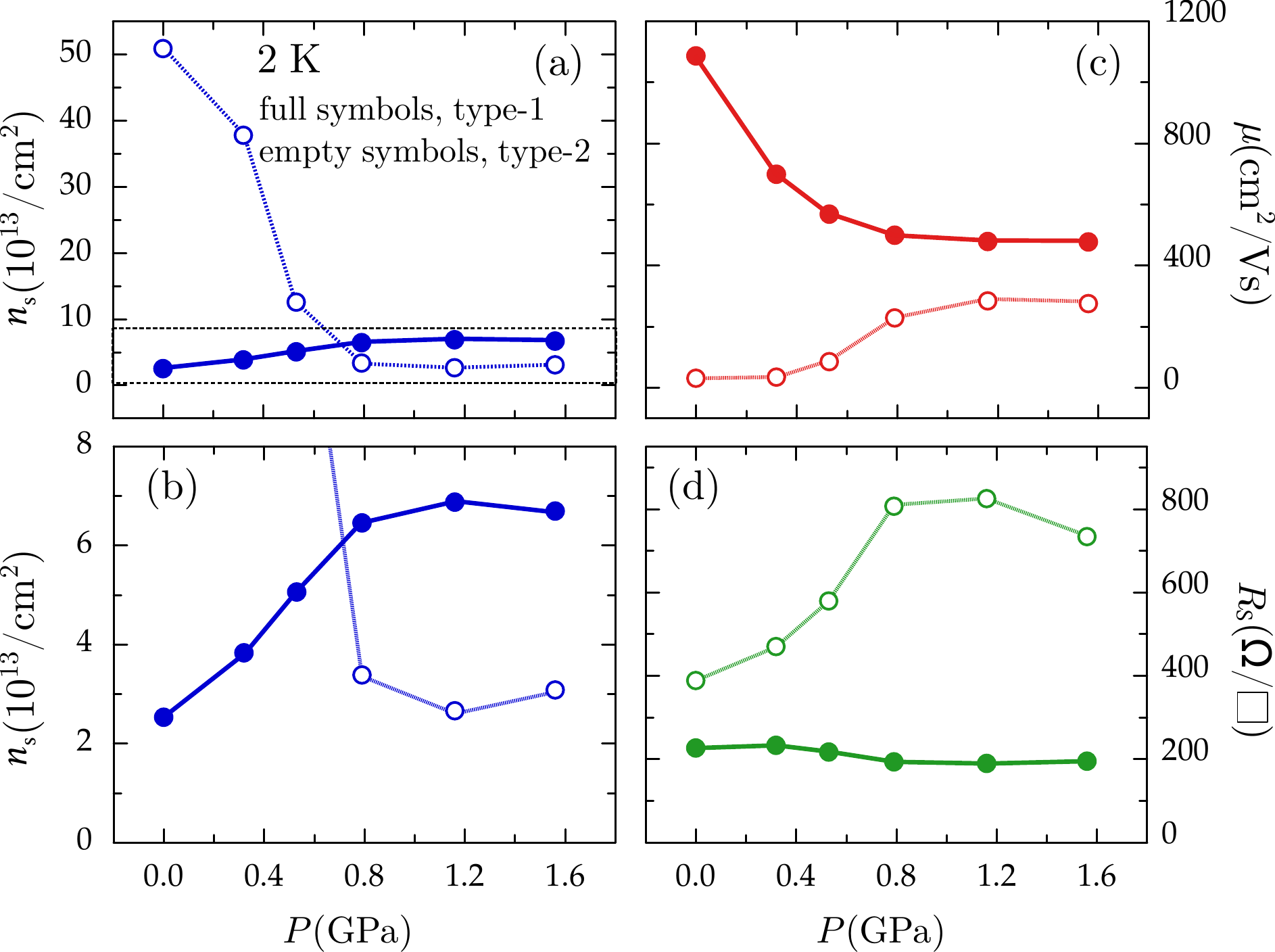}
    \caption{ \label{Fig:ParSampleB} \emph{Multi-band parameters of sample B at 2\,K. Full symbols represent type-1 carriers (electrons) and empty symbols represent type-2 carriers (holes).} (a) 2D carrier density $n_\text{S}$ as a function of pressure. The dashed box indicates the zoomed-in region shown in (b). (c) Mobility $\mu$ and (d) sheet resistance $R_\text{S}$ as a function of pressure for both type of carriers. Lines are a guide to the eye.}
\end{figure}

The fitting parameters for the densities and mobilities are shown in Fig.~\ref{Fig:ParSampleB}.
We find that the 2D carrier density $n_\text{S}$ of the type-1 carriers increases considerably by $\sim$\,170\,$\%$ at 1.13\,GPa (Fig.~\ref{Fig:ParSampleB}(a) and (b), full symbols), whereas the mobility $\mu$ drops to half (Fig.~\ref{Fig:ParSampleB}(c), full symbols). For the type-2 carriers (empty symbols), the effect is opposite, with a dramatic drop of $n_\text{S}$ already at low pressures and an increase in $\mu$ with increasing pressure. The corresponding sheet resistance $R_\text{S}$ of type-1 carriers alone decreases by about 20\,$\%$, whereas that of type-2 carriers doubles. For clarity, we show here the data set for sample B; sample A shows the same trend (see Fig.~\ref{Fig:Par_SamplesAB} of Appendix\,~\ref{app:sup}). 
In conclusion, hydrostatic pressure increases the density of the electronic carriers while suppressing their mobility, as already revealed by a simplified single-band analysis and in agreement with {\it ab initio} predictions.\cite{Zabaleta2016} Nevertheless, the two-band fitting shown here adds new features to the oversimplified one-band scenario: the resistance of the electronic carriers does not increase under pressure, but slightly decreases and, interestingly, a new type of hole-like carrier is revealed. Under increasing pressure, this type-2 carrier behaves approximately opposite to the type-1 electronic carrier.


\section{Discussion}
\label{sec:disc}

Before summarizing the results from the evaluation of the magnetotransport data, it is worthwhile to address two essential issues: The relevance of electron-electron interaction and the possibility of a localization transition (MIT) in the
disorder problem with spin-orbit coupling (symplectic universality class).
 
\subsection{Electron-electron interaction}

Positive MR can have several origins: it can either arise from a multi-band Hall response, or from quantum corrections like WAL or electron-electron interaction (EEI).\cite{Bishop1982} In all three cases, the sheet resistances follow a quadratic behaviour in small magnetic fields, and it may be challenging to decide between these scenarios. 

Recently, Fuchs \textit{et al.}~\cite{Fuchs2015} reported resistance measurements of LAO/STO heterostructures  under pressure. 
Their temperature dependence of the sheet resistance is in good agreement with measurements in our samples (see also Ref.~\onlinecite{Zabaleta2016}). However, they interpret the field-dependence of the magnetoresistance  
in terms of EEI rather than WAL, because the sheet resistance in zero field rises slightly with decreasing temperature, indicating insulating behavior at $T=0$. This seems to be inconsistent with WAL, which requires the system to be on the conducting side of the MIT. We shall show below that this reasoning neglects the multi-band character of the interface.  

First, we like to discuss why the EEI framework offers no potential explanation of our magnetoresistance data. 
EEI as well as multi-band effects lead to MR signals that  grow monotonically with magnetic field.\cite{Bishop1982} In two dimensions, a quadratic increase of the sheet resistance is expected for low magnetic fields, whereas it changes to  logarithmic growth for high fields\cite{Lee1982, Kawabata1981, Lee1985}
\begin{align}
    \Delta \sigma (B, T) &= - \frac{e^2}{h} \frac{\widetilde F_\sigma}{2 \pi}  f \left( \frac{ g \mu_\text{B} B}{ k_\text{B} T} \right) \label{eq:seei} \\[0.3cm]
    f(h) &\approx \begin{cases}
                \ln \left( \frac{h}{1.3} \right)  & h \gg 1 \\[0.2cm]
                0.084 h^2 & h \ll 1 
            \end{cases} \label{eq:seeifine}
\end{align}
and $\widetilde F_\sigma$ is a function of the interaction integrated over the Fermi surface, independent of $B$. 
In the presence of SOC this result is not altered qualitatively.\cite{Millis1984} In contrast to Eqs.~\eqref{eq:seei} and \eqref{eq:seeifine}, the slope of  $\Delta \sigma(B)$ arising from WAL clearly changes sign at a magnetic field $B_c$ [see Fig.~\ref{Fig:Fit}(b)]. 

We have performed our self-consistent data analysis, replacing the WAL formula by  EEI. 
In almost all measurements, we find no meaningful fit. For example,
to fit the curve shown in Fig.~\ref{Fig:Fitcomp}, negative logarithmic behaviour is expected in the magnetoconductivity for $B \gg 1.5\,\text{T}$ due to Eqs.~\eqref{eq:seei} and \eqref{eq:seeifine}, but we find a fit only for a {\it positive} sign of the logarithm, in contradiction to the EEI prediction.
We conclude that EEI cannot explain our experimental results.

In general, magnetotransport measurements in parallel magnetic field may give a more decisive answer to the question of relevant interaction effects. Whereas localization and orbital effects are suppressed in parallel field, interaction effects do not depend on the angle between field and sample. In LAO/STO, however, an unusual MR is found in parallel magnetic field. This negative MR, interpreted recently in Ref.~\onlinecite{Diez2015}, suppresses any possible orbital WAL or EEI contribution and does not allow to distinguish  between different quantum corrections. The parallel field MR for our samples is shown in App.~\ref{app:par} and is in agreement with previous reports.~\cite{Wong2010, Caviglia2010, Shalom2009, Ariando2011, Wang2011, Joshua2013, Sachs2010}


\subsection{Symplectic metal-insulator transition}

As discussed in the previous section, the metallic character in our samples seems to arise from the strong spin-orbit coupling in the system, and not from electron-electron interactions. Therefore, the two-dimensional electronic system in LAO/STO heterostructures may be the ideal candidate to search for the symplectic metal-insulator transition (MIT), predicted theoretically many years ago by Wegner.\cite{Wegner1989} To our knowledge, all reported MITs in two-dimensional systems, first discovered by Kravchenko {\it et al.}~\cite{Kravchenko1994, Kravchenko1995, Kravchenko1996}, have been traced back to  electronic correlations~\cite{Finkelstein2007, Knyazev2008, Kravchenko2010} and may be described with the two-parameter scaling theory by Finkelstein and Punnoose.\cite{Punnoose2001, Punnoose2005} These MITs have been identified only at very low charge carrier densities and correspondingly weak screening of the Coulomb interaction. In contrast, the LAO/STO  interface is a fundamentally different system with relatively high charge carrier density, naturally creating strong screening. The additional presence of considerable spin-orbit coupling in one of the bands furnishes then a two-dimensional system of electrons with no long-range interaction, belonging to the symplectic universality class, for which renormalization group arguments predict a MIT driven by the electron density.\cite{Wegner1989}  
Therefore, hydrostatic pressure and gating might be used as parameters to drive the system through the symplectic metal-insulator transition. However, the presence of several bands with different behavior masks this transition in measurements of the {\it total} conductance.

The conductance $g_1(T)$ of the anti-localized band alone is expected to show one-parameter scaling in the temperature $T$,\cite{Markos2006}
\begin{align}
    \frac{g_1(T)}{g_\text{c}} &=    \exp \left( \tilde{n} \tilde{T}^{-\frac{1}{z \nu}}  \right), \label{eq:scaling} \\
    \tilde{n} &= A_0(n_c - n), \label{eq:scaling2} \\
    \tilde{T} &= \left( \frac{T}{T_0} \right) \label{eq:scaling3}
\end{align}
where $g_\text{c}$ is the critical conductance, $n$ ($n_\text{c}$) the (critical) driving parameter, $\nu \approx 2.75$ is the symplectic critical exponent\cite{Markos2006, Asada2002, Asada2004} and $z$ is the dynamical exponent. The parameters $A_0$ and $T_0$ are determined by microscopic properties of a specific sample. 

\begin{figure}
    \includegraphics[width=\columnwidth]{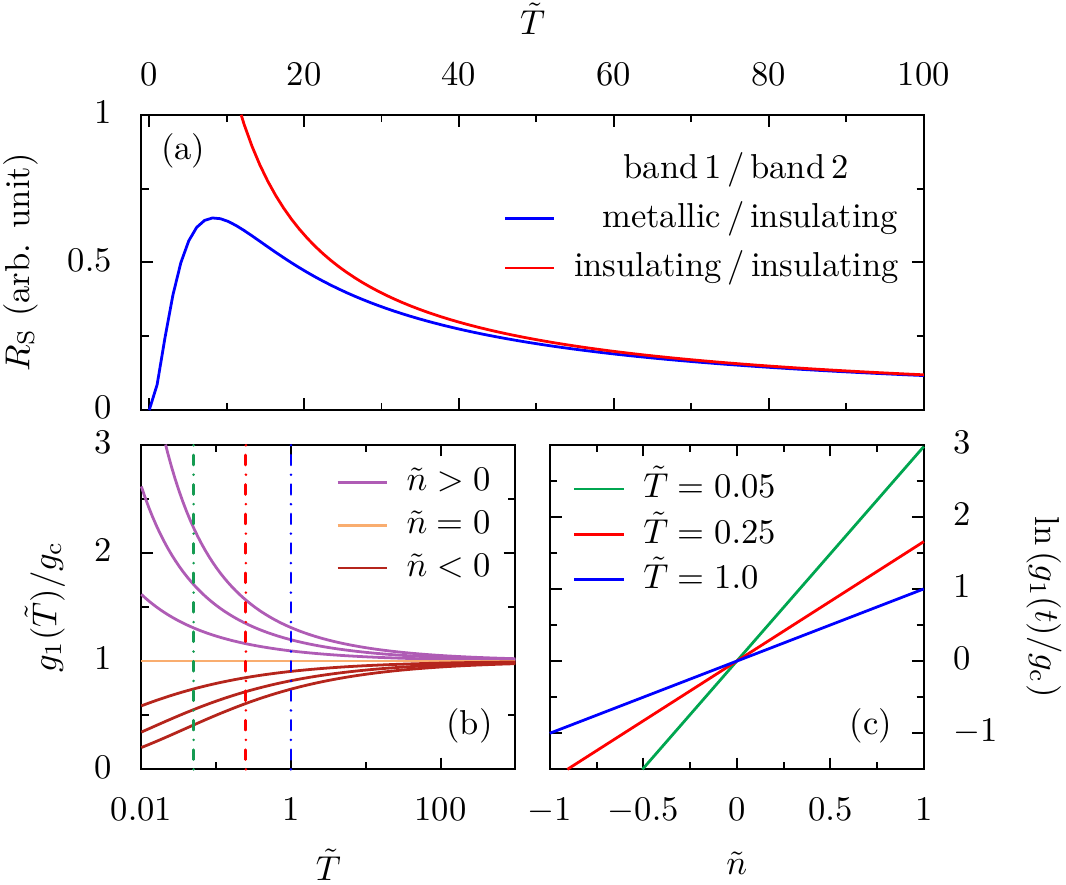}
       \caption{ \label{Fig:scaling} \emph{Scenarios for the sheet resistance and scaling behavior.} (a)\  In a multi-band system, the sheet resistance  $R_\text{S}=1/(g_1+g_2)$ vanishes at $T=0$ on the metallic side of the MIT, although it may show insulating characteristics in an intermediate temperature regime. Exemplarily, we plot $R_\text{S}$ for $g_1(T)$ as specified in Eq.~\eqref{eq:scaling} and for the second band we choose $g_2(T)=aT$ where $a= g_\text{c}\,\text{K}^{-1}$, the dynamical exponent is chosen $z=1$.
(b)\ The expected behavior of the conductance $g_1$ for different densities as a function of the temperature and for different temperatures as a function of the density (c). The crossing point in the right panel indicates the critical density $n_c$ at which the metal-insulator transition takes place at $T=0$. $\tilde{n}$ and $\tilde{T}$ are defined in Eqs.~\eqref{eq:scaling2} and~\eqref{eq:scaling3}.  
}
\end{figure}

For LAO/STO, the sheet resistance is reported to increase with decreasing temperature.\cite{Fuchs2015} This effect appears to be very weak in our samples and
may originate from the second band with a negligible spin-orbit coupling, expected to exhibit localization at low temperatures. The conductance of the two bands add and give a total resistance of $R_\text{S}=(g_1 + g_2)^{-1}$, which ist plotted as a function of the dimensionless temperature~$\tilde{T}$ in Fig.~\ref{Fig:scaling}(a) (blue line).
%
The localized band may dominate the transport for higher temperatures, and the total resistance resembles that of an insulator. However, for low temperatures the scaling in the anti-localized band will finally lead to a vanishing sheet resistance.

Using the multi-band analysis presented in the previous sections, the metallic band featuring the strong SOC may be resolved. Its temperature dependence is then given by~\eqref{eq:scaling}, as plotted in Fig.~\ref{Fig:scaling}(b). 
We believe that the temperature range, which allows for a clear vision of the MIT, was not reached in the current experiments. In two dimensions, the localization length can be quite large, even larger than the sample size. Therefore, the scaling should be visible only for very low temperatures, similar to the interaction-driven transition at low densities. It is also quite probable that the transition will only be revealed under high pressure, rendering future experiments in this direction a challenging task. 


\section{Conclusions}
\label{conclusion}

The evaluation of magnetotransport measurements presents a particular challenge for the LaAlO$_3$/SrTiO$_3$ interface electronic systems. These two-dimensional electron fluids are multi-band systems with strong spin-orbit coupling and quantum interference effects which may result from Anderson localization or electron-electron interaction. Here we investigated interfaces with sufficiently large carrier filling so that a multi-band description is appropriate. This becomes evident from the quadratic terms in the magnetoresistance for intermediate magnetic fields as well as from the cubic term in the Hall resistance. It is then hydrostatic pressure that enhances the visibility of quantum corrections to the magnetotransport, which we conjecture to originate in weak anti-localization. This interpretation is in accordance with the analysis of ambient pressure experiments by Caviglia {\it et al.}~\cite{Caviglia2010} but in discrepancy to the interpretation of Fuchs {\it et al.}~\cite{Fuchs2015} on their pressure dependent data.  

For the interface of LaAlO$_3$/SrTiO$_3$, our work is the first to separate the multi-band behavior from the corrections to magnetotransport using an unbiased self-consistent evaluation procedure. Assuming the corrections to reflect anti-localization, this procedure allows us to distinguish clearly different spin-orbit coupling scenarios. As the interface is subject to a structural inversion-symmetry breaking and strong spin-orbit scatterers are absent, we conclude that the ILP model is the appropriate theoretical framework to analyze the weak anti-localization corrections. We identify a spin-orbit coupling with cubic momentum-dependent spin splitting as the coupling mechanism. Correspondingly, the measured spin-orbit coupling is not Rashba-like but rather ``Dresselhaus-like''. As the Ti $d_{{xz}}$/$d_{{yz}}$ bands display such a cubic spin splitting in a well-established six-band evaluation,~\cite{Zhong2013} we relate the weak anti-localization to the corresponding pair of bands in the electronic structure (see Fig.~\ref{Fig:bands}). A similar correspondence was put forward for SrTiO$_3$ surface states.~\cite{Nakamura2012}

We find an excellent fit over the full measured magnetic field range. However, the nature of the observed second (hole) band remains open. Also it is not clear from the outset, why hydrostatic pressure enhances the weak anti-localization corrections relative to the multi-band signal. We know from previous work that the carrier density in the $d_{{xz}}$/$d_{{yz}}$ bands increases with pressure~\cite{Fuchs2015, Zabaleta2016} and that an {\it ab initio} evaluation is consistent with this observation.~\cite{Zabaleta2016} Yet an explanation of the pressure dependence of the weak anti-localization is beyond the scope of this paper.

We have further shown that electron-electron interactions, invoked by Fuchs {\it et al.}~\cite{Fuchs2015} to explain the logarthmic decrease of the sheet resistance with decresing temperature below 10\,K, do not easily account for our data. Moreover, the rise of the total resistance with decreasing temperature, ruling out weak anti-localization in a single band model, can be explained within our multi-band scenario and therefore does not contradict  weak anti-localization as cause for the corrections. However, further measurements at much lower temperatures are needed to confirm the prediction of metallic  $d_{{xz}}$/$d_{{yz}}$ bands. Very low temperatures are also mandatory to observe the scaling behavior [Fig.~\ref{Fig:scaling}(b)] of the conductance associated to  the electron band, which should undergo a two-dimensional metal-insulator transition as a function of the density, even without appreciable electron-electron interaction. This exotic type of Anderson transition --- never observed before --- is hidden in the LaAlO$_3$/SrTiO$_3$ system due to the presence of the second band, which possesses weak spin-orbit coupling and must therefore localize for low temperatures. Only the disentangling of both bands and their respective transport characteristics using the analysis presented in the present work may reveal the symplectic Anderson transition of the LaAlO$_3$/SrTiO$_3$ interface, adding another unexpected facet to this well-studied oxide heterostructure.


\begin{acknowledgments}
This work was  supported by the DFG through the TRR~80. We gratefully acknowledge  S.~C.~Parks and B.~Baum for technical support and V.~S.~Borisov, D.~Fuchs, P.~J.~Hirschfeld, and R.~Valenti for helpful discussions.
\end{acknowledgments}


\appendix

\section{Specification of the Punnoose formula}
\label{app:Pun}

For completeness, we here specify Eq.~\eqref{EqPunnoose1} as given in Ref.~\onlinecite{Punnoose2006}. The constant $C$ enforces the vanishing of the magnetoconductivity in zero field:
\begin{widetext}
\begin{align}
        C &= -2 \ln \left( 1 + \frac{ B_\text{so} }{ B_\text{i} } \right) - \ln \left( 1 + \frac{2 B_\text{so}}{B_\text{i} } \right) + \frac{ 8 }{ \sqrt{ 7 + 16 \left( \dfrac{B_\text{i} }{B_\text{so} } \right) } } \cos^{-1} \left( \frac{ \left( \dfrac{2 B_\text{i} }{B_\text{so} } \right) - 1 }{ \sqrt{ \left[ \left( \dfrac{2 B_\text{i} }{B_\text{so} } \right) + 3 \right]^2 -1 } } \right) \\
    v_s     &= 2 \delta \cos \left[ \theta - \frac{2 \pi}{3} (1-s) \right] \\
    u_s     &= \frac{ 3 v_s^2 + 4  v_s \left( \dfrac{ B_\text{so} }{B} \right) + 5 \left( \dfrac{ B_\text{so} }{B} \right)^2 + 4 \left( \dfrac{ B_\text{i} }{B} \right) \left( \dfrac{ B_\text{so} }{B} \right) -1 }{ \prod\limits_{s^\prime \neq s} \left( v_s - v_{s^\prime} \right) } \\
    \delta  &= \sqrt{ \frac{ 1 - 4 \left( \dfrac{ B_\text{i} }{B} \right) \left( \dfrac{ B_\text{so} }{B} \right) -  \left( \dfrac{B_\text{so} }{B} \right)^2 }{3} } \\
    \theta  &= \frac{1}{3} \cos^{-1} \left[ - \left( \dfrac{  B_\text{so}  }{B \delta} \right)^3 \left( 1 + \dfrac{2 B_\text{i}}{B_\text{so}} \right) \right].
\end{align}
\end{widetext}

\section{Coupling of WAL and multi-band Hall effect: Full results}
\label{app:Coup}

The resistivity tensor for each of the two bands with index $j$ is given by 
\begin{equation}
    \rho_j (B) = 
    \begin{pmatrix}
        \dfrac{1}{\sigma_{0, j} + \delta \sigma_j} & \dfrac{B}{e n_{j}} \\[0.5cm]
        - \dfrac{B}{e n_{j} } & \dfrac{1}{\sigma_{0, j} + \delta \sigma_j}
    \end{pmatrix},
\end{equation}
where $\sigma_{0, j}$ is the Drude conductivity of the respective band and $\delta \sigma_j$ the quantum correction due to localization (which is magnetic field dependent).
Further, we define
\begin{equation}
    \begin{split}
    \rho_0 &= \frac{1}{ \sigma_{0, 1} + \delta \sigma_1(0) + \sigma_{0, 2} + \delta \sigma_2(0) } \\
           &= \frac{1}{ n_1 \mu_1 + n_2 \mu_2 }, 
    \end{split}
\end{equation}
where $\mu_j$ is the charge carrier mobility and $n_j$ the three-dimensional charge carrier density. Therefore, the part of the localization contribution not depending on the magnetic field, is implicitly contained in the values for the density and mobility. This part cannot be addressed via magnetotransport measurements.
We further define $\Delta \sigma_j = \Delta \sigma_j (B) =  \delta \sigma_j(B) - \delta \sigma_j(0)$ and find for the full resistivity tensor
\begin{widetext}
\begin{align}
    \dfrac{\rho_{xx}(B)}{\rho_0} &= \dfrac{ \dfrac{1}{1 + \rho_0 \left( \Delta \sigma_1 + \Delta \sigma_2 \right)  } + \dfrac{\rho_0 \left( e n_1 \mu_1 + \Delta \sigma_1 \right)  \left( e n_2 \mu_2 + \Delta \sigma_2 \right)  }{ \left[ 1 + \rho_0 \left( \Delta \sigma_1 + \Delta \sigma_2 \right) \right]^2  } \left( \dfrac{ e n_1 \mu_1 + \Delta \sigma_1 }{n_1^2 } + \dfrac{ e n_2 \mu_2 + \Delta \sigma_2 }{ n_2^2 } \right) \dfrac{B^2}{e^2} }{ 1 + \dfrac{\rho^2_0 \left( e n_1 \mu_1 + \Delta \sigma_1 \right)^2 \left( e n_2 \mu_2 + \Delta \sigma_2 \right)^2  }{ \left[ 1 + \rho_0 \left( \Delta \sigma_1 +\Delta \sigma_2 \right) \right]^2  } \left( \dfrac{1}{n_1} + \dfrac{1}{n_2} \right)^2 \dfrac{B^2}{e^2} }, \label{eqrho1} \\[0.5cm]
    \dfrac{\rho_{xy}(B)}{\zeta} &=  \dfrac{  \left( \dfrac{ \left(e n_1 \mu_1 + \Delta \sigma_1 \right)^2 }{n_1} + \dfrac{ \left( e n_2 \mu_2 + \Delta \sigma_2 \right)^2 }{n_2} \right) \dfrac{B}{e} +  \dfrac{ \left( e n_1 \mu_1 + \Delta \sigma_1 \right)^2 \left( e n_2 \mu_2 + \Delta \sigma_2 \right)^2\left( n_1 + n_2 \right)}{n_1^2 n_2^2}  \dfrac{B^3}{e^3}}{ 1 + \dfrac{\rho^2_0 \left( e n_1 \mu_1 + \Delta \sigma_1 \right)^2 \left( e n_2 \mu_2 + \Delta \sigma_2 \right)^2  }{ \left[ 1 + \rho_0 \left( \Delta \sigma_1 +\Delta \sigma_2 \right) \right]^2  } \left( \dfrac{1}{n_1} + \dfrac{1}{n_2} \right)^2 \dfrac{B^2}{e^2} }, \label{eqrho2} \\[0.3cm]
    \zeta &= \dfrac{\rho^2_0}{\left[ 1 + \rho_0 \left( \Delta \sigma_1 + \Delta \sigma_2 \right) \right]^2}. \label{eqrho3}
\end{align}
\end{widetext}
Please note that the magnetic field dependence (as well as the dependence on the effective inelastic and SOC fields, $B_\text{i}$ and $B_\text{so}$) of $\Delta \sigma (B)$ is surpressed for sake of a clearer notation. As it is more convenient experimentally for a two-dimensional system to consider the sheet resistance $R_\text{S}$ rather than the resistivity $\rho$, the densities in the main text are two-dimensional quantities in the sense that $\mu^{-1}=e n_\text{3d} \rho = e n_\text{3d} \lambda \, (\rho/\lambda) = e n_\text{2d} R_\text{S} $, where $\lambda$ is the thickness of the quasi-two-dimensional layer.

\section{Supporting experimental material}
\label{app:sup}

\begin{figure}
    \includegraphics[width=\columnwidth]{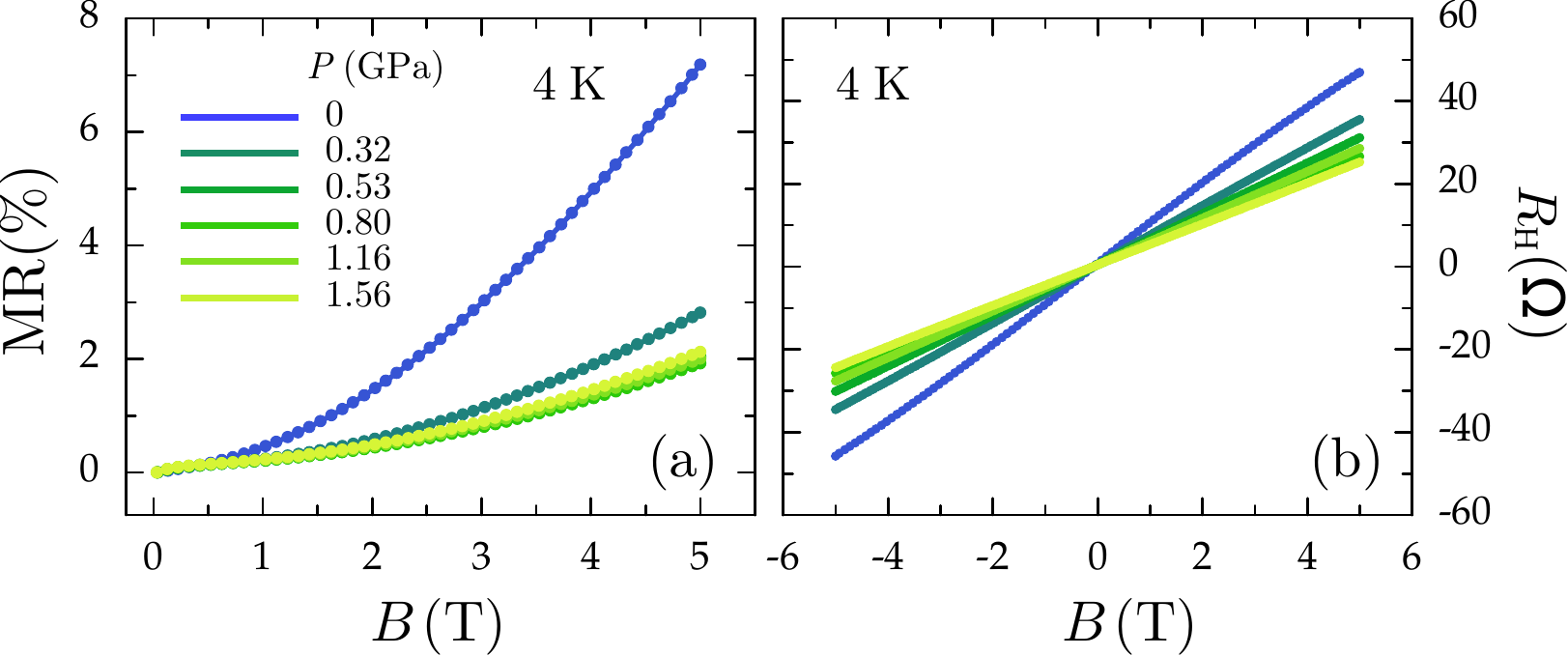}
    \caption{ \label{Fig:MRRH_SampleB} \emph{Experimental magnetotransport results for sample B at 4\,K.} Pressure suppresses the multi-band character as shown by both (a) magnetoresistance MR and (b) Hall resistance $R_\text{H}$ data.}
\end{figure}

\begin{figure}
    \includegraphics[width=\columnwidth]{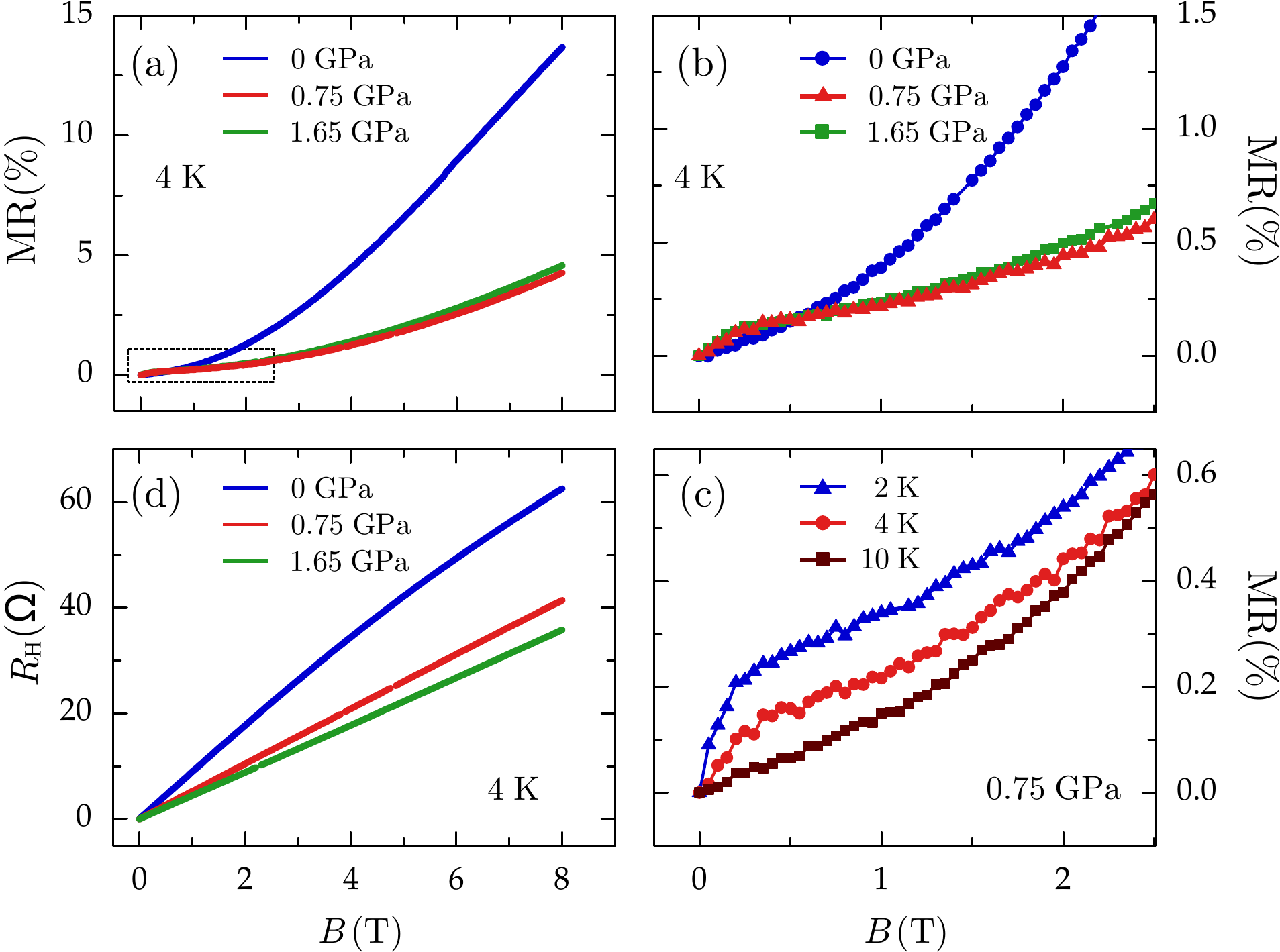}
    \caption{ \label{Fig:MRRH_SampleASymm} \emph{ Magnetotransport results for sample A (shown in Fig.~\ref{Fig:SampleA_MRRH}) after symmetrizing and anti-symmetrizing the sheet resistance $R_\text{S}$ and the Hall resistance $R_\text{H}$ data, respectively}. Transverse components are excluded in the $R_\text{S}$  resistance values by averaging the $R_\text{S}$  measured at identical positive and negative fields. MR is then extracted from the symmetrized $R_\text{S}$ values [(a), (b) and (c)]. (d) The anti-symmetrized $R_\text{H}$ shows the averaged difference between the $R_\text{H}$ values measured at identical positive and negative magnetic fields.}
\end{figure}

\begin{figure}
    \includegraphics[width=\columnwidth]{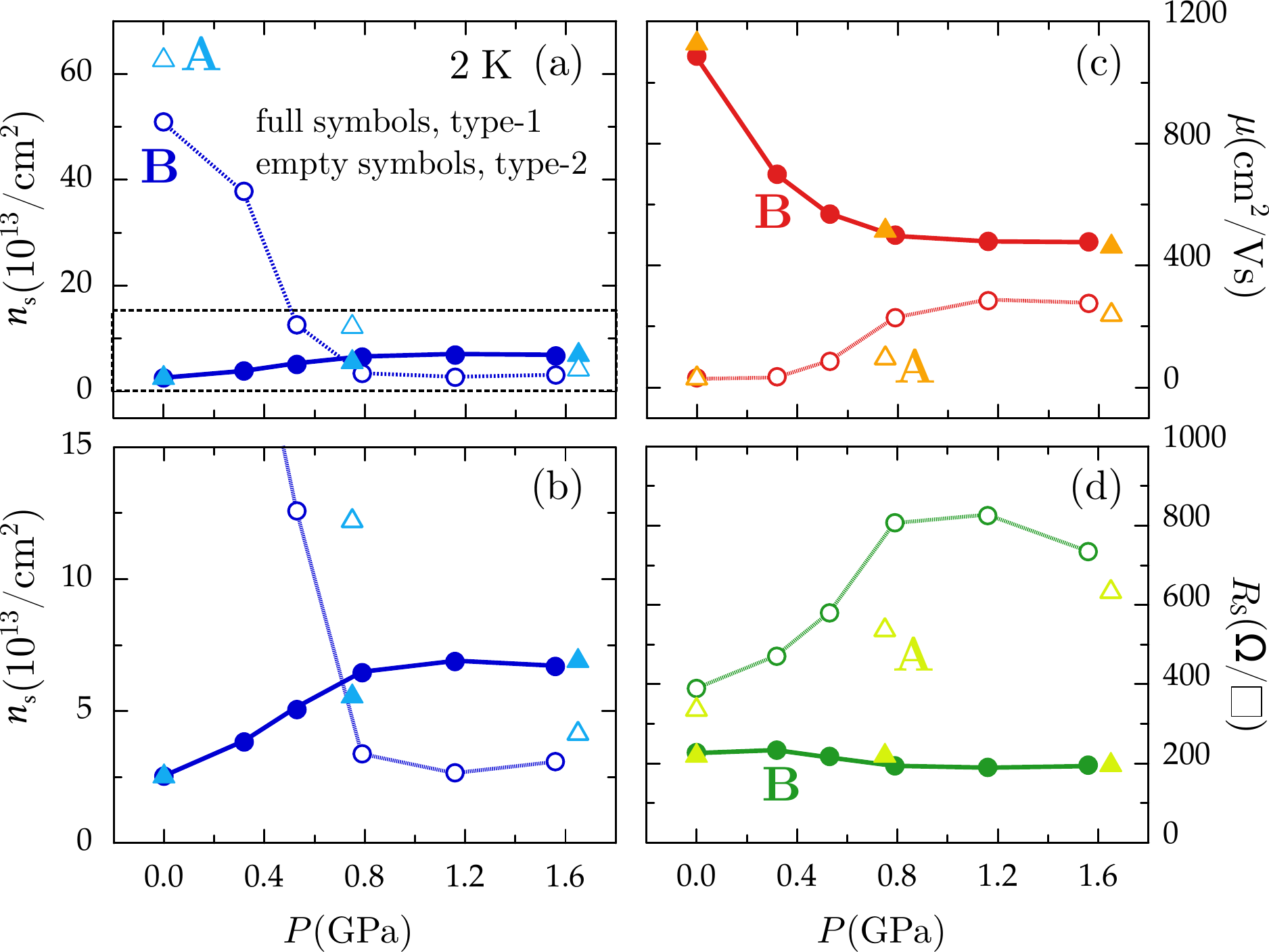}
      \caption{\label{Fig:Par_SamplesAB} \emph{Multi-band parameters of sample A (triangles) and B (circles) at 2\,K.} (a) 2D carrier density $n_\text{S}$ as a function of pressure. The dashed box indicates the zoomed-in region shown in (b). (c) Mobility and (d) sheet resistance evolution with pressure for both type of carriers. Both samples show the same trend. Lines are a guide to the eye.}
\end{figure}

For the sake of completeness, we include here supporting experimental data that show the reproducibility of our results (Fig.~\ref{Fig:MRRH_SampleB} and Fig.~\ref{Fig:Par_SamplesAB}), along with the result of (anti)symmetrizing the magnetotransport data for the subsequent self-consistent fitting procedure described in the main text (Fig.~\ref{Fig:MRRH_SampleASymm}). 

\section{Results in parallel magnetic field}
\label{app:par}

\begin{figure}
    \includegraphics[width=\columnwidth]{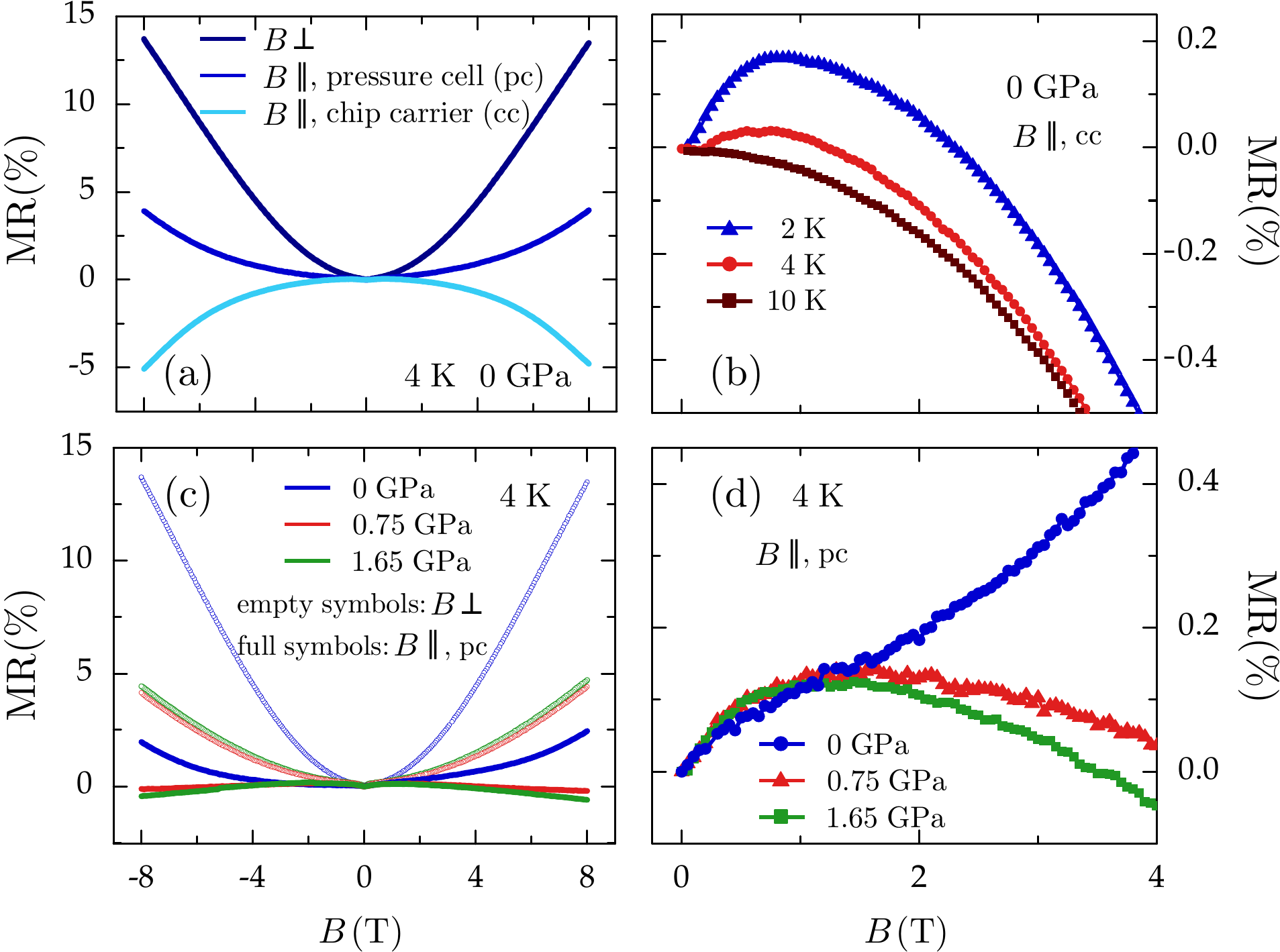}
    \caption{ \label{Fig:MR_Parallel} \emph{Experimental magnetotransport results including measurements with the magnetic field applied parallel to the LAO/STO interface.}(a) Comparison of the MR at 4\,K and 0\,GPa for the magnetic field perpendicular and parallel to the interface (see text). (b) The negative MR shown by the interface under parallel magnetic field develops a positive slope for low fields and low enough temperatures. (c) and (d) Hydrostatic pressure reduces the positive MR for $B$ perpendicular to the interface [see also Fig.~\ref{Fig:SampleA_MRRH}(a) and (b)], whereas it enhances the negative slope observed for $B$ parallel to the interface.}
\end{figure}

Figure~\ref{Fig:MR_Parallel} includes the results of measuring magnetotransport with the magnetic field parallel to the LAO/STO interface. Our pressure setup does not allow for the sample surface to be perfectly parallel to the  magnetic field, nor for the offset to be exactly quantified. On mounting the sample the offset can be estimated to be within $\pm$10{\textdegree} from perfect parallel alignment, whereas on introducing the sample stage in the container with the pressure transmitting fluid it is no longer possible to address the exact offset at which the sample lies during pressure application. Figure~\ref{Fig:MR_Parallel}(a) shows the magnetoresistance curves at 4\,K and 0\,GPa of the LAO/STO interface in the different configurations: under perpendicular magnetic field, with the field parallel to the interface within the pressure cell, and the results of a control experiment with the interface perfectly parallel to the magnetic field (sample mounted on a chip carrier). As seen in the figure, parallel magnetic field produces a negative magnetoresistance that reaches $\sim$-~5\,$\%$ at $\pm$8\,T. Within the pressure cell, the offset of the sample is such that the MR measurement reflects a certain contribution of the perpendicular field component. Figure~\ref{Fig:MR_Parallel}(b) shows a closeup of the MR for the purely parallel field experiment at 0~\,GPa and various temperatures. The 2\,K data clearly reveal the positive slope at low fields which then turns negative for fields higher than $\sim$~1\,T, in agreement with previous reports.~\cite{Wong2010, Caviglia2010, Shalom2009, Ariando2011, Wang2011, Joshua2013, Sachs2010} That feature completely disappears at 10\,K. Interestingly, hydrostatic pressure enhances the negative magnetoresistance, as seen in the comparison between the two magnetic field alignments, perpendicular and parallel, shown in Fig.~\ref{Fig:MR_Parallel}(c) and in the zoomed area plotted in Fig.~\ref{Fig:MR_Parallel}(d).



%


\end{document}